# Experimental Realization of Non-Abelian Permutations in a Three-State Non-Hermitian System

Weiyuan Tang[1], Kun Ding[2], Guancong Ma[1]

[1]Department of Physics, Hong Kong Baptist University, Kowloon Tong, Hong Kong, China

[2]Department of Physics, State Key Laboratory of Surface Physics, and Key Laboratory of Micro and Nano Photonic Structures (Ministry of Education), Fudan University, Shanghai 200438, China

**ABSTRACT**

Eigenstates of a non-Hermitian system exist on complex Riemannian manifolds, with multiple sheets connecting at branch cuts and exceptional points (EPs). These eigenstates can evolve across different sheets, a process that naturally corresponds to state permutation. Here, we report the first experimental realization of non-Abelian permutations in a three-state non-Hermitian system. Our approach relies on the stroboscopic encircling of two different exceptional arcs (EAs), which are smooth trajectories of order-2 EPs appearing from the coalescence of two adjacent states. The non-Abelian characteristics are confirmed by encircling the EAs in opposite sequences. A total of five non-trivial permutations are experimentally realized, which together comprise a non-Abelian group. Our approach provides a reliable way of investigating non-Abelian state permutations and the related exotic winding effects in non-Hermitian systems.

Email: kunding@fudan.edu.cn (D. K.) phgcma@hkbu.edu.hk (G. M.)

**INTRODUCTION**

Permutation is a process of both fundamental and practical importance. For example, one way to distinguish fermions from bosons is to consider the exchange of the wavefunctions of two or more identical particles. Permutations of multiple states can emerge as the phenomenon of multi-state geometric phases [1,2]. They are generally non-commutative and can therefore be mapped to non-Abelian groups. This perspective suggests the possibility of emulating non-Abelian permutations by the parallel transport of three or more degenerate states. However, despite notable attempts in the fields of optics [3], cold atoms [4], and other topological systems [5,6], its realization remains a considerable experimental challenge, with the excitation and manipulation of multiple degenerate but coupled modes posing a major obstacle.

Recent advances in non-Hermitian physics have sparked the development of many intriguing applications related to optics and other classical waves [7,8]. Although non-Hermitian systems can be straightforwardly constructed from Hermitian systems by the inclusion of loss and/or gain, or non-reciprocal hopping, they possess unique characteristics that are not found in their Hermitian counterparts. Perhaps the most notable distinction is that the eigenvalues are generally complex numbers. This simple fact permits the existence of multiple eigenvalue Riemann sheets connected at branch cuts [9–11]. The endpoints of the branch cuts are branch-point singularities known as exceptional points (EPs). Encircling an EP inevitably crosses one or multiple branch cuts, a process that causes the eigenstates to be exchanged and can even produce fractional winding numbers [11–17]. These fascinating behaviors, which are useful for topological energy transfer [11] and asymmetric mode switching [16] applications, have a topological origin: a non-Hermitian system lives on a complex Riemannian manifold that naturally permits state permutations. Hence, non-Hermitian systems offer a new vantage point for the study of state permutations. Recent theoretical investigations suggest that the encircling of multiple order-2 EPs or higher-order EPs is non-Abelian in character [13,18–21] and can give rise to a myriad of exotic winding effects [16,22–25]. These findings are consistent with the group theory point of view, as at least three degrees of freedom are required for non-Abelian processes to emerge. However, experimental confirmations of these proposals are lacking.

In this work, we theoretically investigate and experimentally realize the non-Abelian permutations of three states in a non-Hermitian system. By embedding the system's Riemannian manifolds in a three-dimensional (3D) parameter space, two exceptional arcs (EAs), smooth trajectories of order-2 EPs, are found. As we will show, encircling them induces a unique permutation of the eigenstates. Five distinct types of state permutations are realized by encircling the EAs individually or sequentially. These five permutations, together with an identity element, holistically form a dihedral group of degree three, called the $D_3$ group, which can be used to describe the symmetry operations on an equilateral triangle, as shown in Fig. 1. All five permutations in Fig. 1(a) and the equivalent permutations in Fig. 1(b) are experimentally realized via a stroboscopic approach [17,23,26–28] in acoustic experiments. We further show that the permutation operations are described by $3 \times 3$ unitary matrices, also known as $U(3)$ non-Abelian Berry phases (NABPs), which connect the three-state evolutions on the system's complex Riemannian manifold.

**RESULTS**

**EAs in a three-state non-Hermitian system.** We begin with an exceptional nexus (EX) which emerges in a three-state non-Hermitian Hamiltonian $H = (\omega_0 + i\gamma_0)I + H_{EP}$, where $\omega_0 + i\gamma_0$ denotes the complex onsite energy, and $H_{EP}$ determines the core physics and has the following form

$$H_{EP}(\eta, \zeta, \xi) = \kappa \begin{bmatrix} \sqrt{2}(i+\eta) & 1 & 0 \\ 1 & i\zeta + \xi & 1 \\ 0 & 1 & -\sqrt{2}(i+\eta) \end{bmatrix} + i\sqrt{2}\kappa \begin{pmatrix} g & 0 & 0 \\ 0 & 0 & 0 \\ 0 & 0 & -g \end{pmatrix}. \quad (1)$$

$H_{EP}$ lives on a 3D parameter space spanned by $(\eta, \zeta, \xi) \in \mathbb{R}^3$. There is also another parameter $g$, which for the convenience of discussion is not regarded as a separate dimension. Here, all coefficients are normalized by $\kappa$ (where $\kappa < 0$), which is the hopping coefficient between neighboring sites. A ternary cavity system can be used to experimentally realize the Hamiltonian in acoustics, as shown in Fig. 2(a). The second-order cavity mode is chosen as the onsite resonance mode. The parameters $\eta$ and $\xi$ represent detuning to onsite resonant frequencies, while $i\zeta$ and $ig$ are introduced as losses. Figure 2(b) shows the three eigenmode profiles from a lower frequency (State-1) to a higher frequency (State-3) in the absence of non-Hermiticity. More details of the experimental setup are given in Section III of the Supplementary Information.

When $g = 0$, an EX exists at $(\eta, \zeta, \xi) = (0,0,0)$, which is an order-3 EP that connects to four EAs [23], each of which is a trajectory of order-2 EPs formed by two of the three eigenstates of Eq. (1). These three eigenstates constitute a Hilbert space, which can be figuratively referred to as a fiber, at each parametric point $(\eta, \xi, \zeta)$, thus forming fiber bundles that stick to the base manifold in the parameter space. The non-Hermiticity of the system means that the three eigenvalue Riemannian sheets connect at branch cuts, which naturally allows the exchange of states by encircling the EPs. Hence, each EA can be characterized by the two surrounding eigenstates in permutation. As shown in Fig. 1, the permutations $\mu_1$ and $\mu_3$ constitute two generating operations of the $D_3$ group, and the other elements of the $D_3$ group can be generated by ordered operations of $\mu_1$ and $\mu_3$, i.e., $\rho_1 = \mu_1 \circ \mu_3$, $\rho_2 = \mu_3 \circ \mu_1$, and $\mu_2 = \mu_3 \circ \mu_1 \circ \mu_3$. The identity element is not of interest here, since it generates no changes. The issue of how to realize two EAs that possess the $\mu_1$ and $\mu_3$ types of permutation is therefore crucial for the demonstration of non-Abelian permutations.

In order to achieve this, we introduce the second term in Eq. (1). When $g \neq 0$, the EX splits into two order-2 EPs in the $\zeta\xi$ plane at $\eta = 0$. In this way, the four EAs converging at the EX become a pair of smooth EAs. Figure 2(c) shows the two EAs, denoted $\alpha$ and $\beta$, for $g = 0.61$. We note that the way in which the EAs connect is dependent on the sign of $g$ (see Sections I and II of the Supplementary Information for details), and we focus on the case with positive values of $g$ in the main text. This configuration allows us to trace the evolution of states around the EAs, making it suitable for analyzing the non-Abelian permutation of states that is the focus of this work.

**Two generating permutations by encircling an EA.** We first demonstrate two generating operations, $\mu_1$ and $\mu_3$, that exchange two of the three states. To facilitate the discussion, we order the eigenstates based on the real part of the eigenfrequencies at the starting point of the loop. We set $\eta = 0.33$, which is depicted as a light-green plane in Fig. 2(c). The $\zeta\xi$ plane intersects with both EAs at two EPs, as shown by the red dots on the eigenvalue Riemann surface (real part) in Fig. 2(d). The purple loop encircles EA-$\alpha$, which is formed by the coalescence of state-2 and 3 at $\eta = 0.33$. Hence, one complete cycle must cross the branch cut once, resulting in the swapping of the two states, and consequently, the operation $\mu_1: 123 \rightarrow 132$ is realized. Likewise, it is straightforward to see that $\mu_3$, which encloses EA-$\beta$, exchanges state-1 and 2, i.e., $\mu_3: 123 \rightarrow 213$.

These permutations are experimentally observed via a stroboscopic approach. The parameters of the acoustic system are tuned to the specific values defined by the chosen loop. To achieve this, a Green's function method is used to determine the experimental parameters at each parametric point from the measured pressure response spectra (the details of this process are presented in Sections IV and IX of the Supplementary Information). The complex eigenfrequencies are then obtained by using the above parameters from the Green's function method, and their real parts are plotted as the open circles in Figs. 3(b) and 3(e) for $\mu_1$ and $\mu_3$, respectively. The solid lines in the figure show the theoretical results, and their colors share the same notation as in Fig. 1. The measured eigenvalues are schematically labeled on the Riemann surfaces in Figs. 3(a) and 3(d), which clearly delineate the evolutions associated with $\mu_1$ and $\mu_3$. The salient feature that two states exchange at the branch cuts is clearly seen, and thus Figs. 3(b) and 3(e) agree well with our expectation.

Next, the evolutions of eigenfunctions are also obtained experimentally by measuring the acoustic field profile in all three cavities (see Section V of the Supplementary Information for details). The results indeed show the swapping of eigenfunctions across the branch cut where the real parts of the eigenvalues cross. In Fig. 3(c, f), we plot the representative eigenfunctions at five chosen points along the encircling path, and state exchanges are observed. The results shown in Fig. 3(c) can be taken as an example. We see that at starting point-I, state-2 (shown in green) has a large amplitude at site-A (the middle site). As the system is driven along the $\mu_1$ path, the amplitude at site-A gradually diminishes, while that at site-C increases. At the last two points, state-2 at point-IV smoothly connects to state-3 at point-V, as a direct consequence of crossing the branch cut. Likewise, state-3 at point-IV (shown by the red lines) connects to state-2 at point-V. Meanwhile, state-1 remains almost unchanged throughout the evolution. Upon the completion of one closed cycle, the final outcome is the exchange of state-2 and 3.

We further remark that as the parameters change, the eigenfunctions of the three states also vary. It is therefore crucial to correctly identify how the eigenfunctions evolve along the parametric points, especially in the vicinity of the branch cut where the state exchange takes place. We examine the inner products for all the neighboring states, i.e., $\left|\langle \psi_{i,l+1}^L | \psi_{j,l}^R \rangle\right|^2$, where $|\psi_{j,l}^R\rangle$ is the right eigenfunction of the $j$th state at the parametric point $l$, and $\langle\psi_{i,l+1}^L|$ is the left eigenfunction of the $i$th state at point $l+1$, where $i,j = 1, 2, 3$. The two neighboring eigenfunctions that yield an inner product close to unity are connected by parallel transport [31]. This procedure was performed for all states at all the parametric points presented in our work.

The state permutation induced by $\mu_1$ can be captured by a $U(3)$ NABP [1] (see Section VI of the Supplementary Information for details). Using the eigenvectors of $H_{EP}$ as a basis, the NABP for $\mu_1$ is

$$\boldsymbol{U}_{\mu_1} = \begin{pmatrix} 1 & 0 & 0 \\ 0 & 0 & 1 \\ 0 & 1 & 0 \end{pmatrix}. \qquad (2)$$

From Eq. (2), we can further obtain a multiband Berry phase as

$$\Theta_{\mu_1} = -\text{Im}[\ln(\det \boldsymbol{U}_{\mu_1})] = -\pi. \qquad (3)$$

This phase factor can be observed as a $\pi$-phase difference between state-2 at points-I (shown in green) and V (shown in red) in Fig. 3(c). These results are consistent with the knowledge that an order-2 EP possesses a fractional winding number of 1/2, and the fact that encircling the EP twice restores both states with a Berry phase of $\pi$. The $\mu_3$-induced state permutation can also be seen by tracing the eigenfunction evolutions in Fig. 3(f), and its corresponding NABP is

$$\boldsymbol{U}_{\mu_3} = \begin{pmatrix} 0 & 1 & 0 \\ 1 & 0 & 0 \\ 0 & 0 & 1 \end{pmatrix}, \qquad (4)$$

which also yields a Berry phase of $\Theta_{\mu_3} = -\pi$ from Eq. (3). Although $\Theta_{\mu_1}$ and $\Theta_{\mu_3}$ are the same, the two NABPs $\boldsymbol{U}_{\mu_1}$ and $\boldsymbol{U}_{\mu_3}$ are different, and they do not commute.

**Non-Abelian permutations by sequentially encircling two EAs.** As shown in Fig. 1(a), the $D_3$ group has two elements that describe three-state permutations, denoted as $\rho_1: 123 \rightarrow 231$ and $\rho_2 = 123 \rightarrow 312$. These can be attained by concatenating $\mu_1$ and $\mu_3$ in different orders, i.e., $\rho_1 = \mu_1 \circ \mu_3: 123 \rightarrow 213 \rightarrow 231$ and $\rho_2 = \mu_3 \circ \mu_1: 123 \rightarrow 132 \rightarrow 312$, as shown in Fig. 1(b). The permutation outcomes of $\rho_1$ and $\rho_2$ are clearly different, and this is a manifestation of the non-Abelian characteristics, i.e., $\mu_3 \circ \mu_1 \neq \mu_1 \circ \mu_3$.

The three-state permutations are achieved by sequentially encircling both EAs-$\alpha$ and $\beta$. Without loss of generality, we can anchor the two loops $\mu_1$ and $\mu_3$ at a common vantage point $\mathcal{P}(\eta, \zeta, \xi) = (0.33, 0, 0)$, as depicted by the black hexagon in Fig. 2(c). The point $\mathcal{P}$ is also the starting and end point of the encircling. In Fig. 4(a–c), the $\mu_3$ operation is executed first by encircling EA-$\beta$, which swaps state-1 and 2. The $\mu_1$ operation is then carried out by encircling EA-$\alpha$, thus exchanging the new state-2 and 3. The net result is the swapping of all three states, as defined by $\rho_1$. The $\rho_2$ operation is also experimentally achieved by first encircling $\alpha$ and then $\beta$, as shown in Fig. 4(d–f). The two experimental outcomes, i.e., the mode profiles at the parametric point VII in Figs. 4(c) and 4(f), are clearly distinct, thus unambiguously validating the non-Abelian characteristics. Again, we can summarize the three-state permutations with the NABPs as

$$\boldsymbol{U}_{\rho_1} = \boldsymbol{U}_{\mu_1}\boldsymbol{U}_{\mu_3} = \begin{pmatrix} 0 & 1 & 0 \\ 0 & 0 & 1 \\ 1 & 0 & 0 \end{pmatrix}, \quad \boldsymbol{U}_{\rho_2} = \boldsymbol{U}_{\mu_3}\boldsymbol{U}_{\mu_1} = \begin{pmatrix} 0 & 0 & 1 \\ 1 & 0 & 0 \\ 0 & 1 & 0 \end{pmatrix}. \tag{5}$$

It is clear that $\boldsymbol{U}_{\rho_1} \neq \boldsymbol{U}_{\rho_2}$, although the Berry phases in both cases are $\Theta_{\rho_1} = \Theta_{\rho_2} = 0$ [mod($2\pi$)], as can be identified from the mode profiles at the parametric points I and VII in Figs. 4(c) and 4(f). This verifies the non-Abelian character by encircling different types of EAs.

**Multiple permutations of encircling EAs.** We have already demonstrated four of the five non-trivial permutations depicted in Fig. 1(a). The remaining operation is $\mu_2$: 123 → 321, which exchanges states-1 and 3. Following the rule used to label the EAs, we would expect that EPs exist near to the EX that correspond to the permutation $\mu_2$. This can be attained by shifting the encircling loop of EA-$\alpha$ to $\eta = 0$, as shown in Fig. 2(c). At first sight, it seems counterintuitive that $\mu_2$ can exist in our system, since the hopping between site-B and C is zero in Eq. (1). To see how $\mu_2$ emerges, we first note that $\eta$ represents onsite detuning in sites -B and C, and thus letting $\eta$ cross zero causes the inversion of the lowest and highest frequency modes (state-1 and -3). At $\eta = 0$, the two order-2 EPs (EP-$\alpha$ and $\beta$ in Fig. 2e) are linked by a branch cut that is parallel to the $\zeta$ axis, which connects the lowest and highest frequency sheets. Hence, an evolution that follows the blue loop in Fig. 2(e) exchanges states 1 and 3 and leaves state 2 unchanged, thus realizing $\mu_2$.

The $\mu_2$ operation is also experimentally realized using our acoustic system. The results for the eigenvalues and eigenfunctions are shown in Figs. 3(h) and 3(i), respectively, where the exchange of state-1 and 3 can clearly be seen. We have also computed the corresponding NABP

$$U_{\mu_2} = \begin{pmatrix} 0 & 0 & 1 \\ 0 & 1 & 0 \\ 1 & 0 & 0 \end{pmatrix}, \quad (6)$$

and $\Theta_{\mu_2} = -\pi$. We further remark that, as an element in $D_3$, $\mu_2 = \mu_3 \circ \mu_1 \circ \mu_3$ (or $\mu_2 = \mu_1 \circ \mu_3 \circ \mu_1$). This indicates that the permutation $\mu_2$ can be treated as the operation of encircling the EAs multiple times in our non-Hermitian system. To show this, we can shift the position of the blue loop in Fig. 2(c) slightly to $\eta = 0.055$, so that it transverses three different branch cuts, with each traversal exchanging two states. These results are presented in Section VII of the Supplementary Information. Since $\mu_2$ completes the $D_3$ group here, all other operations that encircle the EAs in Fig. 2(c) multiple times must be equivalent to the single operation shown in Figs. 3 and 4.

**DISCUSSION AND CONCLUSION**

A common practice for characterizing topological manifolds is to consider equivalence classes of loops, in which winding numbers play a vital role. Non-Hermitian topology can be characterized by the eigenvalue winding number, sometimes called the eigenvalue vorticity or discriminant number [28,32,33], which is often considered to be sufficient to reveal the topological structure of the complex Riemann surfaces. However, our results show that eigenvalues are not directly associated with state permutations. Even the eigenvector winding numbers underlain by the Berry phase $\Theta$ do not contain explicit information on state permutations. The state permutations and their non-Abelian characteristics are disclosed either by tracing the parallel transport of all three states or by computing the NABP matrix. Hence, the EAs and their interactions constitute the non-Hermitian counterparts of the knot and link structures of nodal lines in Hermitian band structures [19,20,34].

A question naturally arises as to how the winding numbers relate to the non-Abelian permutations demonstrated in this work. To illustrate this, we recall that the two processes defined by $\rho_1$ and $\rho_2$ yield identical Berry phases $\Theta_{\rho_1} = \Theta_{\rho_2} = 0 \,[\mathrm{mod}(2\pi)]$, which can be regarded as the same eigenvector winding number. We have numerically confirmed that the eigenvalue winding numbers for $\rho_1$ and $\rho_2$ are also identical in these two cases, and are consistent with their Berry phases. As discussed above, the evolutions $\rho_1$ and $\rho_2$ are equivalent to performing both $\mu_1$ and $\mu_3$ in opposite orders. However, the two concatenated loops $\mu_1$ and $\mu_3$ are equivalent to the larger loop encircling both EA-$\alpha$ and $\beta$ (see Section VIII of the Supplementary Information). When this loop is followed, three complete cycles are needed to restore all three states, which gives rise to a fractional winding number of 2/3 [23,35]. In other words, one complete parametric cycle following $\rho_1$ and $\rho_2$ does not recover all the states. It follows that the states after one cycle are dependent on the states at the starting point. This is the reason for the non-Abelian outcomes demonstrated in our work.

In summary, we have successfully demonstrated that all the non-trivial operations comprising the $D_3$ group can be realized by encircling EAs in a three-state non-Hermitian system. Our work builds on recent developments in non-Hermitian physics that have introduced a kaleidoscope of EP structures with distinct topological characteristics. Experimentally, our studies are based on the stroboscopic approach so that the non-adiabatic transitions typically encountered in dynamic evolutions can be avoided [16,36–38]. Our work and the methodology can be extended to study knot and link structures formed by different EAs [39–42]. The combined strength of these theoretical developments and experimental techniques in non-Hermitian physics, in conjunction with the rich arsenal of non-Abelian theories, will open new avenues to the discovery of exotic phenomena and the development of rich applications in a diversity of fields. For example, non-Abelian permutations around multiple EAs provide additional degrees of freedom to manipulate wave propagation [16] and on-chip energy transfer [11]. Relating to our work are several recent studies proposing a new class of anyonic-parity-time symmetric systems [48,49] that can benefit applications such as lasers [50]. On the other hand, the existence and evolutions of multiple EPs in a multi-parameter phase space give rise to rich opportunities of more sophisticated usage of EPs, which may benefit applications such as sensors[43,44], absorbers [45,46], scattering control[47], etc.

**FUNDING**

This work was supported by the Hong Kong Research Grants Council (12302420, 12300419, 22302718, C6013-18G), the National Natural Science Foundation of China (11922416, 11802256, 1217040429), and Hong Kong Baptist University (RC-SGT2/18-19/SCI/006).

## AUTHOR CONTRIBUTIONS

D. K. and G. M. conceived the research. T. W. performed the experiments and carried out numerical calculations. All authors developed the theory, analyzed the data, and wrote the manuscript. G. M. led the research.

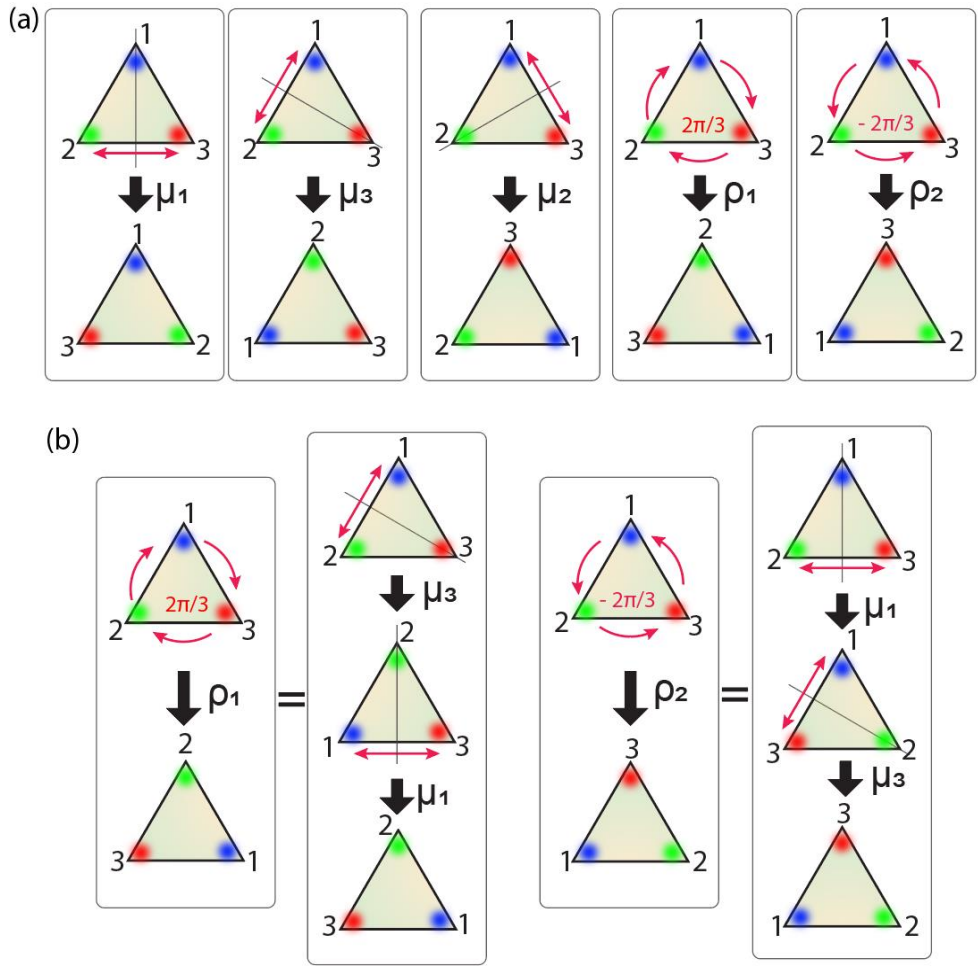

**Figure 1. Operations comprising the $D_3$ group.** (a) The five non-trivial operations of the non-Abelian $D_3$ group depicted as symmetry operations on an equilateral triangle. The $\mu_1$, $\mu_2$, and $\mu_3$ operations flip the triangle about the mirror axis that goes through corners 1, 2, and 3, respectively. The $\rho_1$ and $\rho_2$ operations are the clockwise and anticlockwise rotations of $2\pi/3$ that permutate all three corners. (b) The $\rho_1$ and $\rho_2$ operations can be achieved by concatenating $\mu_1$ and $\mu_3$ in opposite sequences. The non-Abelian nature is clearly seen as $\mu_1 \circ \mu_3 \neq \mu_3 \circ \mu_1$.

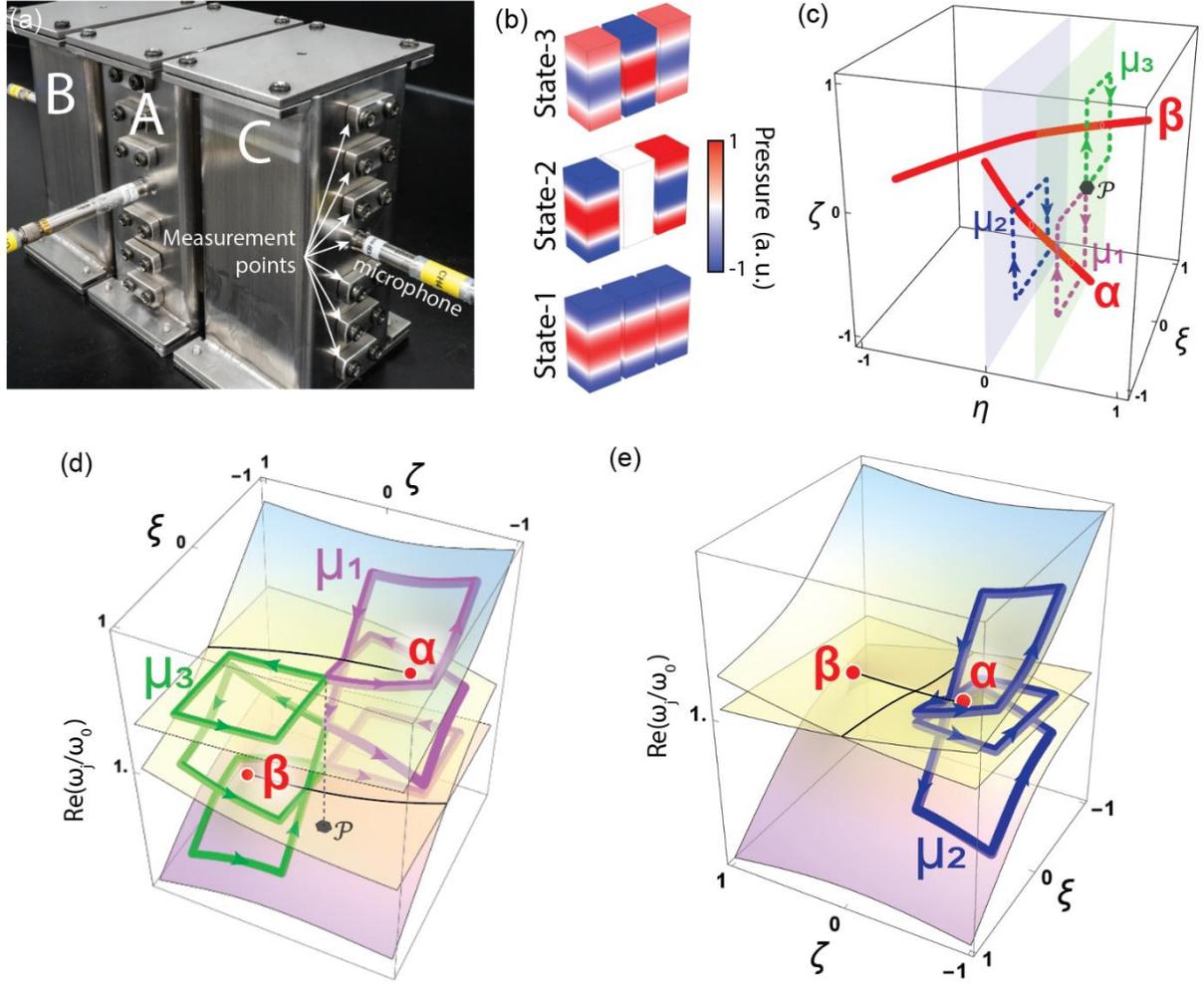

**Figure 2. Three-state acoustic system and state permutations by encircling EAs.** (a) The experimental setup of the ternary coupled acoustic cavities. (b) The simulated acoustic modes in the absence of non-Hermicity ($\eta = \zeta = \xi = g = 0$). (c) Two EAs (solid red curve) lie in a 3D parameter space spanned by $\eta\zeta\xi$. The evolutions along the purple, green, and blue dashed loops produce the operations $\mu_1$, $\mu_3$, and $\mu_2$, respectively. (d) and (e) respectively show the eigenvalue Riemann surfaces on the $\zeta\xi$-plane at $\eta = 0.33$ (the light green plane in c) and $\eta = 0$ (the light blue plane in c). The surfaces from the bottom to top correspond to states-1, 2, and 3 when non-Hermiticity is present. The red dots mark the intersection with the EAs $\alpha$ and $\beta$. The thin black curves are branch cuts, while the purple, green, and blue routes indicate the evolutions of the eigenvalues along $\mu_1$, $\mu_3$, and $\mu_2$, respectively. All eigenvalues are normalized by the onsite resonant frequency $\omega_0 = 19729$ rad/s. The surface hues in (d) and (e) are for aesthetic purposes only, and do not convey physical information.

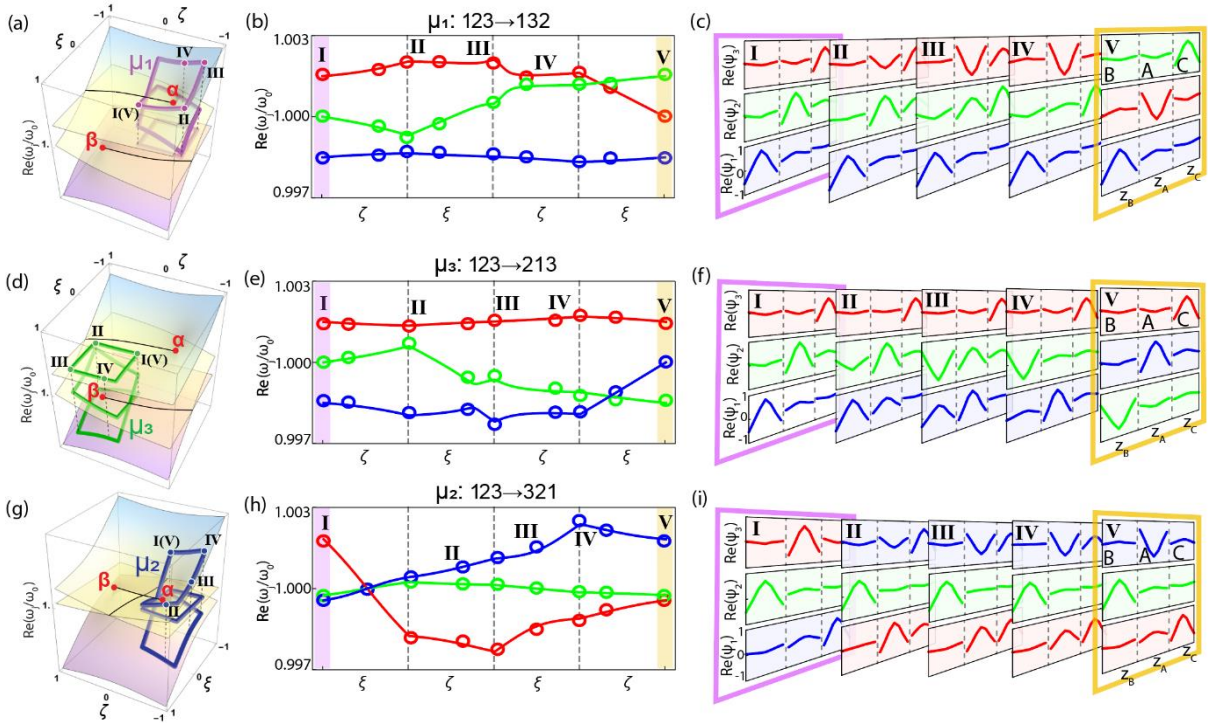

**Figure 3. Two-state permutations.** The three types of two-state permutations are represented by the eigenvalue Riemann surfaces (real parts) in (a) $\mu_1$, (d) $\mu_3$, and (g) $\mu_2$. The corresponding evolutions of the eigenvalues and the measured eigenfunctions are shown in (b, e, h) and (c, f, i), respectively. In (b, e, h), the markers and lines show the experimental and theoretical results, respectively. The eigenvalues and eigenfunctions of states-1, 2, and 3 are labeled in blue, green, and red, respectively. The Roman letters indicate the selected parametric points, which are also labeled in (a, d, g) for better visualization of the encircling evolutions.

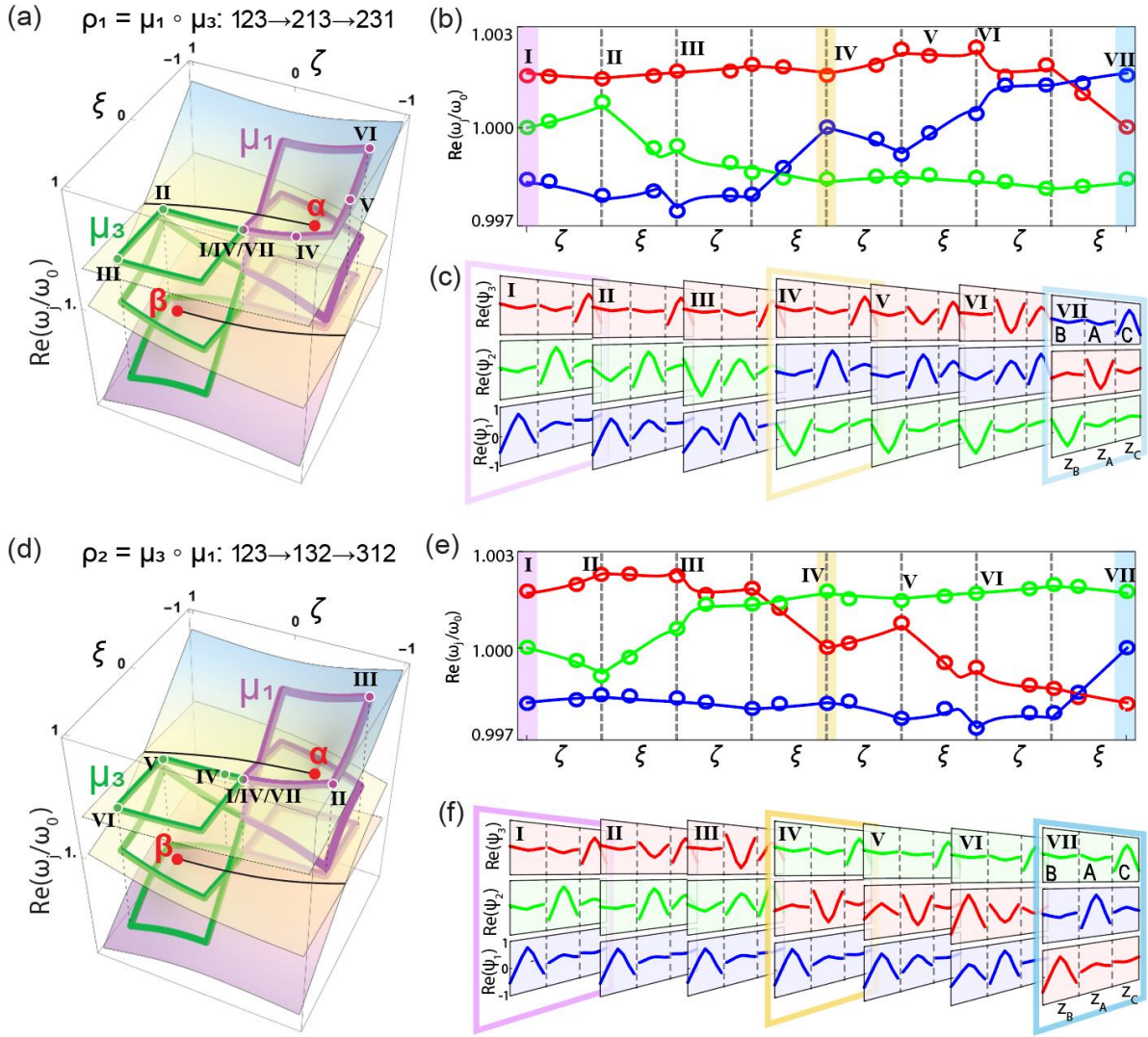

**Figure 4. Three-state permutations and their non-Abelian characteristics.** The two types of three-state permutations $\rho_1$ and $\rho_2$ are represented by their eigenvalue Riemann surfaces (real parts) in (a) and (d), respectively. (b) and (c) respectively show the measured evolutions of eigenvalues and eigenfunctions by encircling first EA-$\beta$ and then EA-$\alpha$, which corresponds to the operation $\rho_1 = \mu_1 \circ \mu_3$. (e, f) The measured evolutions of the eigenvalues and eigenfunctions realize $\rho_2 = \mu_3 \circ \mu_1$. The Roman letters indicate the selected parametric points, which are also labeled in (a, d) for better visualization of the encircling evolutions.

# Supplementary Information

# "Experimental Realization of Non-Abelian Permutations in a Three-State Non-Hermitian System"

Weiyuan Tang[1], Kun Ding[2], Guancong Ma[1]

[1]Department of Physics, Hong Kong Baptist University, Kowloon Tong, Hong Kong, China

[2]Department of Physics, State Key Laboratory of Surface Physics, and Key Laboratory of Micro and Nano Photonic Structures (Ministry of Education), Fudan University, Shanghai 200438, China

This Supplementary Information contains 9 sections, 6 figures, 3 tables, and 7 references.

## I. The exceptional arcs

Here we mathematically show the existence and forms of the exceptional arcs (EAs). We begin with Eq. (1) in the main text, reproduced here

$$H_{EP}(\eta, \zeta, \xi) = \kappa \begin{bmatrix} \sqrt{2}(i+\eta) & 1 & 0 \\ 1 & i\zeta + \xi & 1 \\ 0 & 1 & -\sqrt{2}(i+\eta) \end{bmatrix} + i\sqrt{2}\kappa \begin{pmatrix} g & 0 & 0 \\ 0 & 0 & 0 \\ 0 & 0 & -g \end{pmatrix}, \quad (1)$$

$\kappa, \eta, \zeta, \xi, g \in \mathbb{R}$. Here we set $\kappa = -1$, then the characteristic polynomial $p = \det(\omega \mathbf{I} - H_{EP})$ is

$$p(\omega) = \omega^3 + \omega^2(\xi + i\zeta) - 2\omega(\eta + ig)[\eta + i(2+g)] - 2(\xi + i\zeta)[\eta + i(1+g)]^2. \quad (2)$$

For convenience, we define

$$a_3 = 1, a_2 = (\xi + i\zeta), a_1 = -2(\eta + ig)[\eta + i(2+g)],$$
$$a_0 = -2(\xi + i\zeta)[\eta + i(1+g)]^2. \quad (3)$$

So that Eq. (2) becomes

$$p(\omega) = a_3\omega^3 + a_2\omega^2 + a_1\omega + a_0. \quad (4)$$

Differentiate $p(\omega)$ with respect to $\omega$ produces

$$q(\omega) = b_2\omega^2 + b_1\omega + b_0, \quad (5)$$

where $b_2 = 3a_3, b_2 = 2a_2, b_0 = a_1$. Then, the discriminant $\Delta$ of the characteristic polynomial $p$ is

$$\Delta(p) = \prod_{i<j}(\omega_i - \omega_j)^2 = (-1)^{N(N-1)/2} \det[\mathrm{Syl}(p, q)], \quad (6)$$

where $N = 3$ for our system, and $\mathrm{Syl}(p, q)$ is the Sylvester matrix of the polynomials $p$ and $q$,

$$\mathrm{Syl}(p, q) = \begin{pmatrix} a_3 & a_2 & a_1 & a_0 & 0 \\ 0 & a_3 & a_2 & a_1 & a_0 \\ b_2 & b_1 & b_0 & 0 & 0 \\ 0 & b_2 & b_1 & b_0 & 0 \\ 0 & 0 & b_2 & b_1 & b_0 \end{pmatrix}. \quad (7)$$

Since we only focus on the region in which all four parameters are smaller than 1, we retain the terms up to the third order

$$\Delta(p) \approx -72\xi^2\zeta - 144\xi\eta\zeta - 27\xi^2 + 27\zeta^2 + 192\eta^2 g + 72\zeta^2 g - 64g^3 + i(72\xi^2\eta - 64\eta^3 - 144\xi\zeta g - 72\zeta^2\eta - 54\xi\zeta + 192\eta g^2). \quad (8)$$

At the exceptional points, both the real and imaginary parts of the discriminant $\Delta(p)$ are nil, namely

$$\mathrm{Re}[\Delta(p)] = 0, \mathrm{Im}[\Delta(p)] = 0. \quad (9)$$

In the main text, $g = 0.61$ for the EAs in Fig. 2(c), Eq. (9) gives

$$-72\xi^2\zeta - 144\xi\eta\zeta - 27\xi^2 + 27\zeta^2 + 177.12\eta^2 + 70.92\zeta^2 - 14.53 = 0, \quad (10)$$
$$-64\eta^3 + 72\xi^2\eta - 72\zeta^2\eta - 195.84\xi\zeta + 71.44\eta = 0. \quad (11)$$

Equations (10) and (11) give two sets of curvilinear surfaces in $\eta\zeta\xi$ space that intersect in the formation of the EAs, as depicted in Fig. S1.

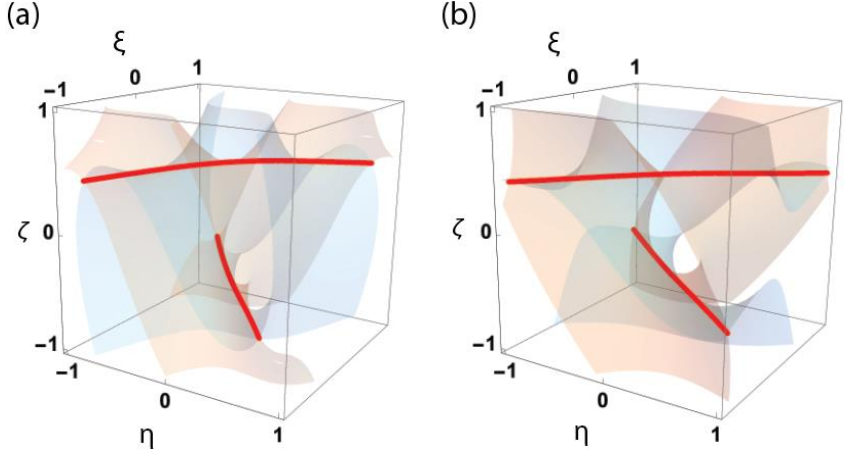

**Figure S1.** (a) The accurate EAs with $g = 0.61$. The orange and blue surfaces correspond to $\text{Re}[\Delta(p)] = 0$ and $\text{Im}[\Delta(p)] = 0$, respectively, and their intersections are highlighted in red, corresponding to the EAs. (b) The EAs obtained by the approximated equations (10) and (11).

## II.  The EAs and the exceptional nexus

The non-Hermitian Hamiltonian, namely Eq. (1) in the main text, is derived from the model used in Ref. [1], which can produce an "exceptional nexus" (EX), i.e., an order-3 EP at which all order-2 EAs form the cusp singularities. When $g = 0$, the EX appears at $\eta, \zeta, \xi = 0$, as shown in Fig. S2(b). Two cusp singularities are perpendicular to each other. Figures S2(a) and S2(c) show the behavior of these EAs when $g < 0$ and $g > 0$, respectively. We see that the EX disappears, and the two cusp singularities formed by the EAs become two smooth EAs. Such configurations are convenient for our investigation of state permutations, since they can be easily achieved and monitored by encircling one EA at a time. Comparing Figs. S2(a) and (c), we can see that the EAs for the $g < 0$ and $g > 0$ cases connect in different manners. When $g < 0$, each EA corresponds to a definite state permutation, but it is not the case when $g > 0$. The state permutation along the EAs can change when $g > 0$, and thus provide the possibility to realize the various permutations in the $D_3$ group.

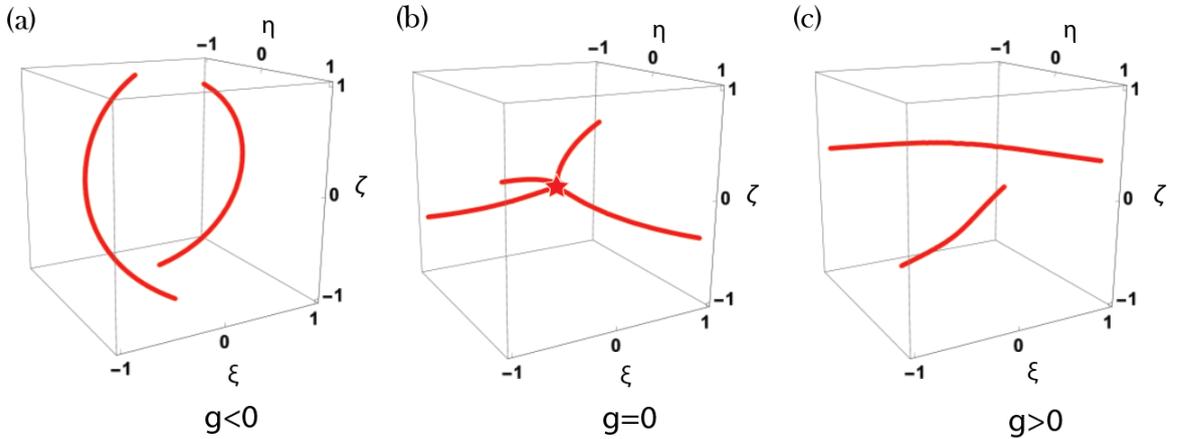

**Figure S2.** The EAs for (a) $g < 0$, (b) $g = 0$, and (c) $g > 0$. When $g = 0$, the EAs kiss at $\eta, \zeta, \xi = 0$ in the formation of an EX, which is an order-3 EP (the red star).

## III. Experimental setup

We use three cuboid acoustic-cavity resonators to realize the non-Hermitian Hamiltonian, as shown in Fig. 2(a) of the main text. The stainless-steel cavities are filled with air and have a height $h = 110$ mm and a square cross-section with a side length of 44 mm. The cavities are joined together by small horizontal holes with a cross-sectional area of 17 mm$^2$, which introduces a hopping of $\kappa = -49.5$ rad/s. The second-order mode, which has a cosine acoustic profile with two nodal planes, is employed to realize the onsite orbital. The Hermitian eigenmodes, that is, with the absence of differential loss and gain, are depicted in Fig. 2(b) in the main text. A small port with a radius of 2 mm is opened on the top of each cavity for the external excitation, and a loudspeaker is used to pump at cavity-B. These ports also introduce additional radiative loss, which contributes to $\gamma_0$ in our model.

The realization of state permutations requires the additional loss and frequency offset to be precisely controlled by the relevant acoustic parameters in our experiment. In our experiments, the additional loss is achieved by placing small pieces of acoustic sponge at the bottom of specific cavities. The frequency offset is achieved by tuning the volume of the acoustic cavity, which is implemented by inserting a specific amount of putty. We have experimentally characterized the effects of sponge and putty, as shown in Fig. S3(a) and (b), respectively. The loss and detuning are determined by fitting the spectral responses of a single cavity using the Green's function.

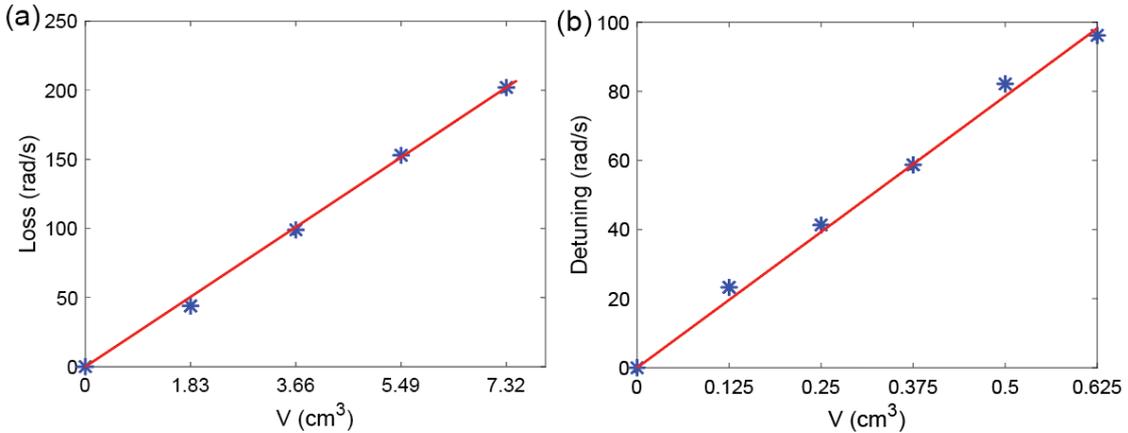

**Figure S3.** (a) The linear relationship between loss and the volume of the acoustic sponge. (b) The linear relationship between detuning and the volume of putty. The red curves are the linear fit, and the blue markers represent extracted data from measurements.

## IV. The acquisition of eigenvalues and other system parameters from experimental data

The eigenvalues of our system can be acquired from the measured pressure response spectra through the Green's function method[1–4]. The Green's function can be expanded in the left and right eigenvectors $|\psi_j^R(\lambda_l)\rangle$ and $\langle\psi_j^L(\lambda_l)|$ with $j = 1, 2, 3$ labeling the states and $\lambda_l$ denoting the parametric coordinate of the $l$-th step along a closed loop

$$\overleftrightarrow{G}(\omega, \lambda_l) = \sum_{j=1}^{3} \frac{|\psi_j^R(\lambda_l)\rangle\langle\psi_j^L(\lambda_l)|}{\omega - \omega_j(\lambda_l)}. \tag{12}$$

Here $\omega_j(\lambda_l)$ is the eigenvalues. The pressure responses measured at a specific parametric step $\lambda_l$ inside the three coupled cavities are

$$P(\omega, \lambda_l) = \langle m|\overleftrightarrow{G}(\omega, \lambda_l)|s\rangle, \tag{13}$$

wherein $|s\rangle$ and $|m\rangle$ are 3×1 column vectors denoting the source and probe positions. In our experiment, the source is placed on the top of cavity A and three identical microphones pick up the pressure response at all three cavities. Therefore, $|s\rangle = (0 \quad 1 \quad 0)^T$ and $|m\rangle = (1 \quad 0 \quad 0)^T, (0 \quad 1 \quad 0)^T$ and $(0 \quad 0 \quad 1)^T$ for the probing at cavity B, A, and C, respectively. The measured data at the three cavities are then fitted against Eq. (13) by a genetic-algorithm-assisted least-square fitting. All the parameters $\omega_0, \gamma_0, \kappa, \eta, \zeta, \xi, g$, together with the eigenvalues $\omega_j$, are thus obtained.

## V. The acquisition of eigenfunctions from experimental data

The permutations of states are observed by tracing the eigenfunctions of the three states as they evolve along different loops. Hence the acquisition of eigenfunctions is a crucial step. For a three-state non-Hermitian system, the eigenfunctions can be constructed by the onsite modes

$$|\psi_j^R(\lambda_l)\rangle = \begin{bmatrix} a_{j,A}(\lambda_l)|\varphi_A\rangle \\ a_{j,B}(\lambda_l)|\varphi_B\rangle \\ a_{j,C}(\lambda_l)|\varphi_C\rangle \end{bmatrix}, \tag{14}$$

$$\langle\psi_j^L(\lambda_l)| = [b_{j,A}(\lambda_l)\langle\varphi_A|, \quad b_{j,B}(\lambda_l)\langle\varphi_B|, \quad b_{j,C}(\lambda_l)\langle\varphi_C|]. \tag{15}$$

Here, $|\varphi_{A,B,C}\rangle$ is the isolated onsite mode of the individual cavity A, B, and C. Its real-space representation is a $7 \times 1$ column vector since there are seven measurement positions on each cavity. Thus, in the real-space representation, $|\psi_j^R(\lambda_l)\rangle$ is a $21 \times 1$ column vector and $\langle\psi_j^L(\lambda_l)|$ is a $1 \times 21$ row vector. The real parts of the 21 elements of $|\psi_j^R(\lambda_l)\rangle$ are the results shown in Fig. 3(c, f, i) and Fig. 4(c, f) in the main text.

To obtain those results, the first step is to obtain $|\varphi_{A,B,C}\rangle$ by the Green's function method mentioned before

$$P_B(\omega) = \frac{\langle m|\varphi_B\rangle\langle\varphi_B|s\rangle}{\omega-[\omega_0+i\gamma_0-i\sqrt{2}\kappa(1+g+\eta)]}, \quad (16)$$

$$P_A(\omega) = \frac{\langle m|\varphi_A\rangle\langle\varphi_A|s\rangle}{\omega-[\omega_0+i\gamma_0+\kappa(i\zeta+\xi)]}, \quad (17)$$

$$P_C(\omega) = \frac{\langle m|\varphi_C\rangle\langle\varphi_C|s\rangle}{\omega-[\omega_0+i\gamma_0+i\sqrt{2}\kappa(1+g+\eta)]}, \quad (18)$$

wherein $|m\rangle$ and $|s\rangle$ now become $7\times 1$ column vectors. $|s\rangle$ only has one nonzero element. The retrieved parameters are used in Eqs. (16-18). The data to be fitted are the measured pressure responses at 31 frequencies near $\omega_0$ at 7 positions on each isolated cavity.

The second step is to obtain the coefficients $a_{j;A,B,C}, b_{j;A,B,C}$. This is done by fitting the pressure responses of the three coupled cavities measured at totally 21 positions (7 for each cavity) at the same 31 frequencies

$$P_j(\omega,\lambda_l) = \langle m|\vec{G}(\omega,\lambda_l)|s\rangle = \sum_{j=1}^{3}\frac{\langle m|\psi_j^R(\lambda_l)\rangle\langle\psi_j^L(\lambda_l)|s\rangle}{\omega-\omega_j}. \quad (19)$$

Here, $|m\rangle$ and $|s\rangle$ are $21\times 1$ column vectors and there is also only one nonzero element in $|s\rangle$. Upon attainment of the coefficients $a_{j;A,B,C}, b_{j;A,B,C}$, the right and left eigenfunctions $|\psi_j^R(\lambda_l)\rangle$ and $\langle\psi_j^L(\lambda_l)|$ are readily obtained. This procedure is repeated for each parametric step $\lambda_l = (\eta,\zeta,\xi)$ along the designated loops.

Because the non-Hermitian system lives on a self-intersecting complex Riemannian manifold, special care must be taken to correctly identify the evolution of eigenstates. First, the parallel transport of states must be satisfied. Our experimental raw data, which are obtained using a stroboscopic approach, inevitably carry arbitrary phases that are caused by the acoustic excitation at each parameter point. The arbitrary phases are extracted as $\theta_j(\lambda_{l+1}) = \text{Im}[\ln\langle\psi_j^L(\lambda_l)|\psi_j^R(\lambda_{l+1})\rangle]$, and then compensated at each step[4,5]. This way ensures that the eigenfunctions at neighboring steps satisfy the parallel transport under a constant $U(1)$ gauge, which is the phase factor at the initial step $\theta_j(\lambda_1)$. Second, to obtain the correct connection of states at neighboring parametric points, the inner products $|\langle\psi_j^L(\lambda_l)|\psi_j^R(\lambda_{l+1})\rangle|$ are computed for all states at all steps as an indicator. This procedure is necessary to identify state exchanges. In addition, the eigenvectors are intrinsically mixed due to the presence of non-Hermicity, meaning that the eigenfunction profiles do not stay the same during the encircling process. The evolution of states is therefore correctly enforced by comparing the inner products at each step.

## VI. The non-Abelian Berry phase matrix

The evolutions of states around one or multiple EPs, including their permutations, can be captured by the non-Abelian Berry phase matrix. For a single band, the Berry phase is a $U(1)$ connection between the eigenstate at different parametric locations in an adiabatic evolution. The formalism can be generalized for a multiband system, in which case the Berry phase becomes a $U(n)$ matrix, wherein $n$ is the number of consecutive bands under consideration [6]. For our system, $n=3$ so that

$$\left|\psi_j^R(\lambda_{\mathcal{L}})\right\rangle = \sum_{k=1}^{n=3} U_{jk} \left|\psi_k^R(\lambda_1)\right\rangle, \tag{20}$$

with $U_{jk}$ is an element in

$$\widetilde{U} = \prod_{l=1}^{\mathcal{L}-1} M(\lambda_l, \lambda_{l+1}), \tag{21}$$

wherein $M_{jk}(\lambda_l, \lambda_{l+1}) = \left\langle \psi_j^L(\lambda_l) \middle| \psi_k^R(\lambda_{l+1}) \right\rangle$. $\widetilde{U}$ is a unitary matrix, but in general, it does not take the forms of $U$ as shown in Eqs. (2, 4, 5, 6) in the main text. To obtain those specific results, the state vectors at the starting point needs to be prepared as $\left|\overline{\psi_1^R}(\lambda_1)\right\rangle = (1 \quad 0 \quad 0)^T$, $\left|\overline{\psi_2^R}(\lambda_1)\right\rangle = (0 \quad 1 \quad 0)^T$, $\left|\overline{\psi_3^R}(\lambda_1)\right\rangle = (0 \quad 0 \quad 1)^T$. (For these three state vectors to be valid, the encircling path must be sufficiently distant from the EP, otherwise the eigenvectors become skewed. This condition is always met in our calculations and experiments, since we do not approach the EPs.) Although it is difficult to actually prepare these state vectors in stroboscopic experiments, they are connected to the eigenvectors by a unitary transformation on the eigenvectors $\left|\overline{\psi_j^R}(\lambda_1)\right\rangle = P \left|\psi_j^R(\lambda_1)\right\rangle$, wherein $P$ is given by $\widetilde{H}_{EP3} = P^\dagger H_{EP3} P$ such that $\widetilde{H}_{EP3}(\lambda_1) \left|\overline{\psi_j^R}(\lambda_1)\right\rangle = \omega_j(\lambda_1) \left|\overline{\psi_j^R}(\lambda_1)\right\rangle$. By applying the same transformation $P$ to the eigenvectors $\left|\psi_j^R(\lambda_l)\right\rangle$ (and $\left\langle \overline{\psi_j^L}(\lambda_l) \right| = P^\dagger \left\langle \psi_j^L(\lambda_l) \right|$) for the subsequent steps $\lambda_l$, we can then obtain the $U$ shown in the main text. It is easy to see that $U$ and $\widetilde{U}$ are connected by the same transformation $U = P^\dagger \widetilde{U} P$.

When the path is a closed loop, i.e., $\lambda_{\mathcal{L}} = \lambda_1$, $U$ is gauge-invariant. We can further obtain a phase factor, sometimes also called a multiband Berry phase

$$\Theta = -\mathrm{Im}[\ln(\det U)]. \tag{22}$$

We remark that the phase factor $\Theta$ given by Eq. (22) is identical to the result obtained by tracing the cyclic evolution of a single eigenstate along an EP-encircling loop multiple times until the recovery of all states. The latter method was used to obtain the non-Hermitian Berry phase, such as in Refs. [1,2,5].

## VII. The permutation of state-1 and 3

In the main text, we show that state-1 and 3 can exchange by encircling an EA in the $\zeta\xi$ plane at $\eta = 0$, which generates the $\mu_2$ operation. By referring to the Cayley table of the $D_3$ group, it is easy to see that $\mu_2 = \mu_3 \circ \mu_1 \circ \mu_3$, which is graphically shown in Fig. S4(a). This process can be found in our system. By setting $\eta = 0.055$, i.e., slightly shifting the light-blue plane and the blue loop in Fig. 2(c) in the main text, the evolution delineates the path shown in Fig. S4(b). The state exchanges take place sequentially, as shown in Fig. S4(c). By shifting back to $\eta = 0$, the three exchanges occur at the same point, which are the results shown in Fig. 3(g-i) in the main text.

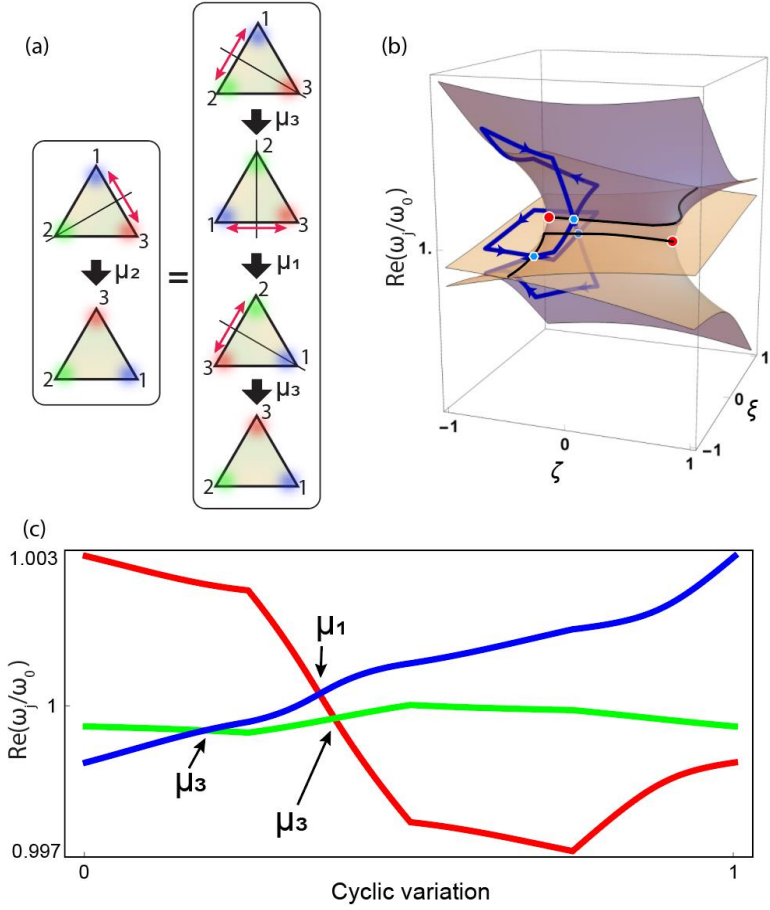

**Figure S4.** (a) The $\mu_2$ operation can be generated by $\mu_2 = \mu_3 \circ \mu_1 \circ \mu_3$. (b) The eigenvalue Riemann surfaces near a pair of EPs (red dots) on the $\zeta\xi$-plane with $\eta = 0.055$. The branch cuts are depicted in black. The blue loop traverses the branch cuts three times and the intersecting points are marked by the blue dots. (c) Unwrapping the evolution, we can clearly identify the composition of $\mu_3 \circ \mu_1 \circ \mu_3$.

## VIII. The equivalence loop of the concatenated loops

The operations $\rho_1$ or $\rho_2$ are generated by executing $\mu_1$ and $\mu_3$ in different sequences. The concatenated loops are equivalent to a single loop encircling both EA-$\alpha$ and $\beta$, as shown in Fig. S5(a). Following this loop, the evolution traverses all three sheets of the Riemannian surface [Fig. S5(b)]. In this case, three complete cycles are needed to recover all the eigenstates [1,7].

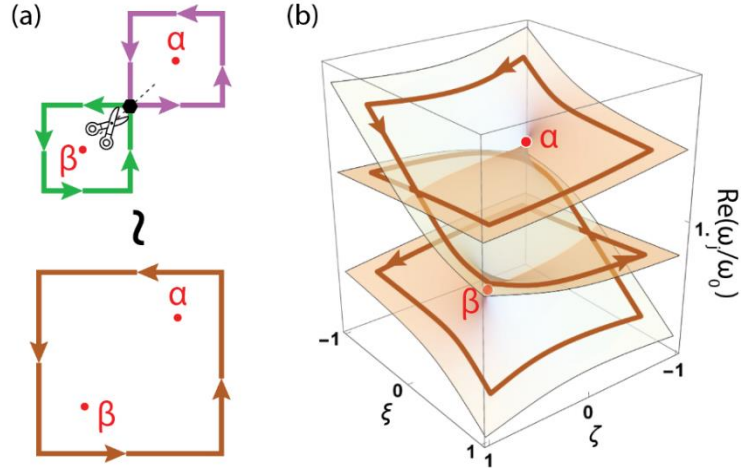

**Figure S5.** (a) The equivalence of the two loops each encircling one EA (also see Fig. 2c, main text) and a big loop that encircles both EAs. (b) The eigenvalue (real parts) Riemann surface shows the encircling of both EA-$\alpha$ and $\beta$ within one loop.

## IX.    Parameters retrieved from the measurements

Here, we present the parameters in our experiments. The second longitudinal mode resonates at $f_0 = 3140$ Hz so that $\omega_0 = 19729$ rad/s. The intrinsic loss of each cavity is $\gamma_0 = 83.5$ rad/s. The parametric points along the $\rho_1$ loop (which includes $\mu_1$ and $\mu_3$), the $\rho_2$ loop and the $\mu_2$ loop are given in Tables S1–S3 accordingly. All these parameters are obtained using the Green's function method as described in Supplementary Information, Section III. To show the validity of our fitting method, we also represent some of the fitting results in Fig. S6.

**Table S1.** The parameters for the $\rho_1$ loop at $\eta = 0.33$.

| Point # | $\zeta$ | $\xi$ |
|---|---|---|
| 1 (I) | 0.00 | 0.00 |
| 2 | 0.16 | 0.00 |
| 3 (II) | 0.54 | 0.00 |
| 4 | 0.54 | 0.35 |
| 5 (III) | 0.54 | 0.51 |
| 6 | 0.16 | 0.51 |
| 7 | 0.00 | 0.50 |
| 8 | 0.00 | 0.30 |
| 9 (IV) | 0.00 | 0.00 |
| 10 | -0.40 | 0.00 |
| 11 | -0.60 | 0.00 |
| 12(V) | -0.60 | -0.16 |
| 13(VI) | -0.60 | -0.44 |
| 14 | -0.36 | -0.41 |
| 15 | 0.00 | -0.46 |
| 16 | 0.00 | -0.26 |
| 17(VII) | 0.00 | -0.00 |

**Table S2.** The parameters for the $\rho_2$ loop at $\eta = 0.33$.

| Point # | $\zeta$ | $\xi$ |
|---|---|---|
| 1 (I) | 0.00 | 0.00 |
| 2 | -0.40 | 0.00 |
| 3 (II) | -0.60 | 0.00 |
| 4 | -0.60 | -0.16 |
| 5 (III) | -0.60 | -0.44 |
| 6 | -0.37 | -0.42 |
| 7 | 0.00 | -0.43 |
| 8 | 0.00 | -0.27 |
| 9 (IV) | 0.00 | 0.00 |
| 10 | 0.16 | 0.00 |
| 11(V) | 0.55 | 0.00 |
| 12 | 0.55 | 0.29 |
| 13(VI) | 0.55 | 0.51 |
| 14 | 0.16 | 0.50 |
| 15 | 0.00 | 0.50 |
| 16 | 0.00 | 0.33 |
| 17(VII) | 0.00 | 0.00 |

**Table S3.** The parameters for $\mu_2$ loop $\eta = 0$.

| Point # | $\zeta$ | $\xi$ |
|---|---|---|
| 1 (I) | -0.22 | -0.46 |
| 2 | -0.20 | 0.00 |
| 3 | -0.21 | 0.44 |
| 4 (II) | -0.57 | 0.40 |
| 5 | -0.79 | 0.40 |
| 6 (III) | -0.79 | 0.00 |
| 7 (IV) | -0.81 | -0.44 |
| 8 | -0.61 | -0.45 |
| 9 (V) | -0.22 | -0.46 |

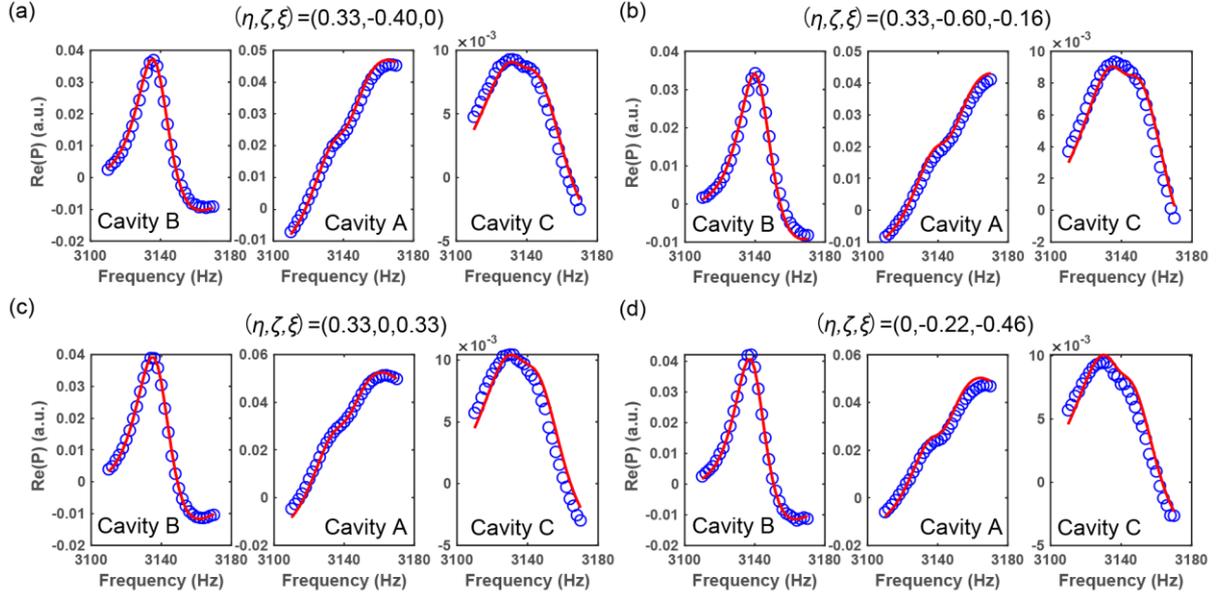

**Figure S6.** Selected results of measured pressure response spectra and fitting results. The blue markers are experimentally measured data. The red curves are fitted by using the Green's function method. Excellent agreement is seen.